\documentclass[a4paper,12pt]{article}
\usepackage{epsfig}
\usepackage{color}
\usepackage[numbers,sort&compress]{natbib}

\newlength{\dinwidth}
\newlength{\dinmargin}
\begin{document}
\title{  $B_s\to K^{(*)-}K^{(*)+}$, $ K^{(*)-}\pi^{+}$, $ K^{(*)-}\rho^{+}$ decays
in R-parity violating  supersymmetry}
\author{Yuan-Guo Xu$^{1,2}$ , ~Ru-Min Wang$^3$ ~  and  Ya-Dong
Yang$^{1,2}$ \footnote{Corresponding author, yangyd@iopp.ccnu.edu.cn}
 \\
 { \small\it $^1$Key Laboratory of Quark $\&$ Lepton Physics (Huazhong Normal
University),} \\{\small\it Ministry of Education, P.R. China }
 \\
 {\small \it  $^2$ Institute of Particle Physics,
 Huazhong Normal University,  Wuhan, Hubei 430079, P.R.China \footnote{Mailing address}}
 \\
{\small \it $^3$Department of Physics, Yonsei University,
Seoul 120-479, Korea}
}

\maketitle
\begin{abstract}
With  the  first measurements of the branching ratios and the direct
CP asymmetries of $B_s\to K^{-}K^{+}$, $K^{-}\pi^{+}$ decays by the CDF collaboration,
we constrain the relevant parameter space of the Minimal Supersymmetric Standard Model
with R-parity violation. Using the constrained R-parity violating couplings, we further examine
their possible effects in $B_{s}\to K^{-*}\pi^{+}$, $K^{(*)-}\rho^{+}$ and $K^{(*)\pm}K^{(*)\mp}$ decays.
We find that some branching ratios and CP asymmetries are very sensitive to the R-parity
violating couplings.  The direct longitudinal CP asymmetries of tree-dominated process
$B_s\to K^{*-}\rho^{+}$ could be enlarged to $\sim 70\%$ and the  longitudinal polarizations of $B_s\to
K^{*-}K^{*+}$, $K^{*-}\rho^{+}$ decays could be suppressed  very much by the squark
 exchange couplings. Near future experiments at CERN LHC can test these predictions
 and shrink/reveal the parameter spaces  of RPV SUSY.
\end{abstract}

\vspace{0.5cm} \noindent {\bf PACS Numbers:  12.60.Jv,
  12.15.Ji, 12.38.Bx, 13.25.Hw}

\newpage
\section{Introduction}
In the recent ten years, the successful running of B factories BABAR
and Belle has provided rich experimental data for $B^{\pm}$ and
$B^{0}$, which has confirmed the Kobayashi-Maskawa CP asymmetry
mechanism in the Standard Model (SM) and also shown hints for new
physics (NP).  Among the rich phenomena of B decays, the two-body
charmless decays are the known effective probes of the CP violation
in the SM and are sensitive to potential NP scenarios beyond the SM.
The two body charmless $B_s$ decays will play the same important
role for studying the CP asymmetries (CPA), determining CKM matrix
 elements  and constraining/seraching for
 the indirect  effects of various NP scenarios.

Recently the CDF collaboration at Fermilab Tevatron has made the
first measurement of charmless two-body $B_{s}$ decays
\cite{Abulencia:2006psa,Morello:2006pv,Morello2008,
Aaltonen2008,updated data}
\begin{eqnarray}
&& \mathcal{B}(B_s \rightarrow
K^-K^+)=(24.4\pm1.4\pm3.5)\times10^{-6},\nonumber\\
&&\mathcal{B}(B_s \rightarrow K^-\pi^+)=(5.0\pm0.7\pm0.8)\times10^{-6},\nonumber\\
&&\mathcal{A}_{CP}^{dir}(B_s \rightarrow
K^-\pi^+)=0.39\pm0.15\pm0.08. \label{data}
\end{eqnarray}
The measurement is an important mark of  $B_{s}$ physics, and also
implies that many $B_{s}$ decay modes could be precisely measured
at the coming LHCb.

Compared with  the  theoretical predictions for these quantities
in Refs. \cite{Beneke:2003zv}, \cite{Ali:2007ff} and
\cite{Williamson:2006hb}, based on the QCD factorization approach
(QCDF)  \cite{BBNS}, the perturbative QCD (PQCD)  \cite{PQCD}, and
the soft-collinear effective theory (SCET) \cite{SCET},
respectively,  one would find the experimental measurements of
branching ratios agree with the SM predictions within their large
theoretical uncertainties. However, NP effects would be still
possible to render other observable deviated from the SM
expectation with the branching ratios nearly unaltered
\cite{Baek:2006pb}.

The related decays $B_s\to K^{(*)-}K^{(*)+}$, $K^{(*)-}\pi^{+}$,
$K^{(*)-}\rho^{+}$ have also been extensively studied in the
literature
\cite{Beneke:2003zv,Beneke:2006hg,Ali:2007ff,Williamson:2006hb,Chen:2001sx,Li:2003hea}.
The four decays $B_s\to K^{(*)-}K^{(*)+}$ are governed by the
$\bar{b}\to\bar{s} u\bar{u}$ transition at the quark level, which
are penguin-dominated processes. The tree-dominated decays $B_s\to
K^{(*)-}\pi^{+}$, $ K^{(*)-}\rho^{+}$  are induced by $\bar{b}\to
\bar{u}u\bar{d}$  where the direct  CPA  are expected to be small
in the SM. At present, among many measurements of $B_{u,d}$
decays, several discrepancies with the SM predictions have
appeared in the corresponding penguin-dominated
$\overline{b}\rightarrow \overline{s}q\overline{q}~(q=u,d,s)$
processes and tree-dominated $\overline{b}\rightarrow
\bar{d}q'\bar{q}'~(q'=u,d)$ processes. For example, $B\rightarrow
\pi\pi,\pi K$ puzzles
 \cite{:2008zza,Aubert:2007mj,Abe:2004us,Aubert:2005av,Aubert:2008sb}
and  the large transverse polarization anomaly in $B\rightarrow
\rho K^*,\phi K^*$ decays
\cite{Aubert:2004qb,Zhang:2005iz,Aubert:2006uk}. Although the
discrepancies are not statistically significant, there is an
unifying similarity pointing to NP (for example,
 \cite{Baek:2006ti,Baek:2006pb,Wu:2006ur,Dariescu:2003tx,Chang:2008tf&2006dh,Buras:2004th,London:2004ej,Yang:2005es}).
There could be also potential NP contributions in $B_s\to
K^{(*)-}K^{(*)+}$, $K^{(*)-}\pi^{+}$, $K^{(*)-}\rho^{+}$ decays,
which have been analyzed  with different NP models
 \cite{Baek:2006pb,Baek:2005wx,Fleischer:2007wg,London:2004ej}. The
measurements given in Eq. (\ref{data}) will afford an opportunity to
constrain NP scenarios beyond the SM.

Among the NP models that survived electroweak data, one of the
respectable options is the R-parity violating (RPV) supersymmetry
(SUSY). The possible appearance of the RPV couplings
 \cite{Weinberg:1981wj,SUSY}, which will violate the lepton and baryon number
conservation, has gained full attentions in searching for SUSY
 \cite{Barbier:2004ez,Chemtob:2004xr,Barbier:1998fe,Allanach:1999bf}.
In this work, we will study the $B_s\to K^{(*)-}K^{(*)+}$,
$K^{(*)-}\pi^{+}$ and $K^{(*)-}\rho^{+}$ decays  in the Minimal
Supersymmetric Standard Model (MSSM) with R-parity violation by
employing the QCDF.  The four $B_s\to K^{(*)+}K^{(*)-}$ decays are
all induced at the quark level by $\bar{b}\to \bar{s}u\bar{u}$
process, they involve the same set of RPV coupling constants. The
$B_s\to K^{(*)-}\pi^{+}$, $K^{(*)-}\rho^{+}$ decays are due to
$\overline{b}\to \overline{d}u\overline{u}$ at the quark level,
and they also involve the same set of RPV coupling constants.
Using the latest experimental data and the theoretical parameters,
we have derived  new bounds on the relevant R-parity violating
couplings, which are consistent with the bounds from $B_{u,d}$
decays. With the constrained parameter spaces, we predict the RPV
effects on the other quantities in $B_s\to K^{(*)-}K^{(*)+}$, $
K^{(*)-}\pi^{+}$ and $ K^{(*)-}\rho^{+}$ decays which have not
been measured yet. We find that the R-parity violating effects on
some branching ratios and direct CPA could be large.
 For example, the squark exchange couplings  could enhance the direct
 CP asymmetry
in  the longitudinal polarized mode of $B_s\to K^{*-}\rho^{+}$  to $\sim 73\%$
 and suppress the longitudinal polarization fractions of $B_s\to K^{*-}K^{*+}$ and
 $K^{*-}\rho^{+}$ to $\sim 0.5$. The mixing-induced CPA  are  also found to be
sensitive to the RPV effects.  Therefore, with the ongoing B-physics at Tevatron,
in particular with the onset of the LHC-b experiment,
 we expect a wealth of  $B_s$ decays data and  measurements of
these observables  could restrict or reveal the NP parameter spaces  in the near future.

The paper is arranged as follows.  In Sec. 2, we give the expression
of  the CP averaged branching ratios, the direct CPA, the
mixing-induced CPA
 and the polarization fractions  within the QCDF
approach in $B_s\to K^{(*)-}K^{(*)+}$, $K^{(*)-}\pi^{+}$,
$K^{(*)-}\rho^{+}$ systems, where   the RPV SUSY effects are
included. We also tabulate the theoretical inputs in this section.
Sec. 3 deals with the numerical results. We display the constrained
parameter spaces which satisfy the present experimental data of
$B_s$ decays,
 and then we use the  constrained  parameter spaces to predict the RPV effects on  the
 other observable quantities, which have not been measured yet in
 $B_s\to K^{(*)-}K^{(*)+}$, $K^{(*)-}\pi^{+}$ and $K^{(*)-}\rho^{+}$ decays.
 Sec. 4 contains our summary and conclusion.

\section{The theoretical frame for $B \to M_1M_2$ decays }
\subsection{ The decay amplitudes  in the SM }
  In the SM, the low energy effective Hamiltonian for
  the $\Delta B=1$ transition at the scale $\mu\sim m_{b}$ is given by   \cite{coeff}
 \begin{eqnarray}
 \mathcal{H}^{SM}_{eff}&=&\frac{G_F}{\sqrt{2}}\sum_{p=u, c}
 \lambda^q_p \Biggl(C_1Q_1^p+C_2Q_2^p
 +\sum_{i=3}^{10}C_iQ_i+C_{7\gamma}Q_{7\gamma}
 +C_{8g}Q_{8g} \Biggl)+ \mbox{h.c.},
 \label{HeffSM}
 \end{eqnarray}
here  $\lambda^q_p=V_{pb}V_{pq}^* $ for $b \to q$ transition
$(p\in \{u,c\},q\in \{d,s\})$ and the detailed definition of the
operator base can be found in   \cite{coeff}.

With the weak effective Hamiltonian given by Eq. (\ref{HeffSM}),  one
can  write the decay amplitudes for the general two-body hadronic
 $B\to M_{1}M_{2}$ decays as
\begin{eqnarray}
  \mathcal{A}^{SM}(B\to M_1M_2)&=&\left< M_1M_2|
  {\cal H}^{SM}_{eff}|B \right> \nonumber\\
  &=&\sum_p \sum_i \lambda^q_p
  C_i(\mu)\left<M_1M_2|Q_i(\mu)|B\right>.
  \end{eqnarray}
The essential theoretical difficulty for obtaining the decay
amplitude arises  from the  evaluation of hadronic matrix elements
$\langle M_1M_2|Q_i(\mu)|B\rangle$, for which we will employ the
QCDF \cite{BBNS} throughout  this paper.

The QCDF  \cite{BBNS} allows us to compute the non-factorizable
corrections to the hadronic matrix elements $\langle M_1
M_2|Q_i|B\rangle$ in the heavy quark limit. The factorization
formula reads
\begin{eqnarray}
\langle M_1 M_2|Q_i|B\rangle=\left(F_j^{B\to
M_1}T^I_{ij}*f_{M_2}\Phi_{M_2}+[M_1\leftrightarrow
M_2]\right)+T^{II}_i*f_{B}\Phi_{B}*f_{M_1}\Phi_{M_1}*f_{M_2}\Phi_{M_2},\label{FactorizationFormula}
\end{eqnarray} where $F_j^{B\to M_1}$ is the appropriate form factor, $\Phi_M$ are
leading-twist light-cone distribution amplitudes and the star
products imply an integration over light-cone momentum fractions.
By the above factorization formula,  the complicated hadronic
matrix elements of four-quark operators are reduced to simpler
non-perturbative quantities and calculable hard-scattering kernels
$T^{I,II}$.

Then the  decay amplitude has the form
\begin{eqnarray}
  \mathcal{A}^{SM}(B\to M_1M_2)
  =\sum_p \sum_i \lambda^q_p
  \Biggl\{a^p_i\langle M_2|J_2|0\rangle\langle
  M_1|J_1|B\rangle+b^p_i\langle M_1M_2|J_2|0\rangle\langle
  0|J_1|B\rangle\Biggl\},\label{AMPLITUDE}
  \end{eqnarray}
 where  the effective parameters $a_i^p$   including nonfactorizable corrections at order
of $\alpha_s $. They are calculated from the vertex corrections, the
hard spectator scattering, and the QCD penguin contributions. The
parameters $b^p_i$ are calculated from the weak annihilation
contributions. The factorized matrix
element is given by
\begin{eqnarray}
A_{M_1M_2}\equiv \langle M_2|(\bar{q}_2\gamma_\mu(1-\gamma_5)q_3)
 |0\rangle
 \langle M_1|(\bar{b}\gamma^\mu(1-\gamma_5)q_1)|B \rangle,
 \end{eqnarray}
 which can be expressed in terms of the corresponding decay constants and form
factors.  We will use the QCDF amplitudes of these decays derived in the comprehensive papers
 \cite{Beneke:2003zv,Beneke:2006hg} as  inputs for the SM amplitudes.

\subsection {R-parity violating SUSY effects in the decays}

In the most general superpotential of MSSM, the RPV superpotential
is given by
  \cite{Weinberg:1981wj}
\begin{eqnarray}
\mathcal{W}_{RPV}&=&\mu_i\hat{L}_i\hat{H}_u+\frac{1}{2}
\lambda_{[ij]k}\hat{L}_i\hat{L}_j\hat{E}^c_k+
\lambda'_{ijk}\hat{L}_i\hat{Q}_j\hat{D}^c_k+\frac{1}{2}
\lambda''_{i[jk]}\hat{U}^c_i\hat{D}^c_j\hat{D}^c_k, \label{rpv}
\end{eqnarray}
where $\hat{L}$ and $\hat{Q}$ are the SU(2)-doublet lepton and
quark superfields and $\hat{E}^c$, $\hat{U}^c$ and $\hat{D}^c$ are
the singlet superfields, while $i$, $j$ and $k$ are generation
indices and $c$ denotes a charge conjugate field.

The bilinear RPV superpotential terms $\mu_i\hat{L}_i\hat{H}_u$ can
be rotated away by suitable redefining the lepton and Higgs
superfields   \cite{Barbier:2004ez}. However, the rotation will
generate a soft SUSY breaking bilinear term which would affect our
calculation through penguin level. However, the processes discussed
in this paper could be induced by tree-level RPV couplings, so that
we would neglect sub-leading RPV penguin contributions in this
study.

The $\lambda$ and $\lambda'$ couplings in Eq. (\ref{rpv}) break
the lepton number, while the $\lambda''$ couplings break the
baryon number. There are 27 $\lambda'_{ijk}$ couplings, 9
$\lambda_{ijk}$ and 9 $\lambda''_{ijk}$ couplings.
$\lambda_{[ij]k}$ are antisymmetric with respect to their first
two indices, and $\lambda''_{i[jk]}$ are antisymmetric with $j$
and $k$.

\begin{figure}[htbp]
\begin{center}
\begin{tabular}{c}
\includegraphics[scale=0.7]{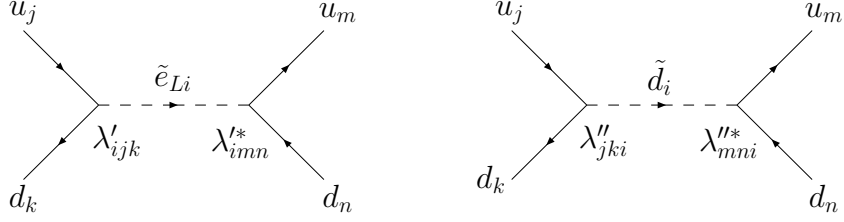}
\end{tabular}
\end{center}
\vspace{-0.6cm} \caption{\small RPV tree level contributions to
$b\to u\bar{u} q$ process.}
 \label{RPVfig}
\end{figure}

From Eq. (\ref{rpv}), we can obtain the relevant RPV effective
Hamiltonian as shown in Fig. \ref{RPVfig}
\begin{eqnarray}
\mathcal{H}^{RPV}_{eff}&=&\sum_i\frac{\lambda'_{ijm}
\lambda'^*_{ikl}}{2m^2_{\tilde{e}_{Li}}}
\eta^{-8/\beta_0}(\bar{d}_m\gamma^\mu P_Rd_l)_8(\bar{u}_k
\gamma_\mu
P_Lu_j)_8\nonumber\\
&&+\sum_n\frac{\lambda''_{ikn}
\lambda''^*_{jln}}{2m^2_{\tilde{d}_{n}}}\eta^{-4/\beta_0}
\left\{\left[(\bar{u}_i\gamma^\mu P_Ru_j)_1(\bar{d}_k\gamma_\mu
P_Rd_l)_1-(\bar{u}_i\gamma^\mu P_Ru_j)_8(\bar{d}_k\gamma_\mu
P_Rd_l)_8\right]\right. \nonumber
\\&&-\left[\left.(\bar{d}_k\gamma^\mu P_Ru_j)_1
(\bar{u}_i\gamma_\mu P_Rd_l)_1 -(\bar{d}_k\gamma^\mu P_Ru_j)_8
(\bar{u}_i\gamma_\mu P_Rd_l)_8\right]\right\}, \label{EqRPVH}
\end{eqnarray}
where Eq. (\ref{EqRPVH}), $P_L=\frac{1-\gamma_5}{2}$,
$P_R=\frac{1+\gamma_5}{2}$,
$\eta=\frac{\alpha_s(m_{\tilde{f}})}{\alpha_s(m_b)}$ and
$\beta_0=11-\frac{2}{3}n_f$. The subscript of  the currents
$(j_{\mu})_{1, 8} $ represents  the current in the color singlet
and octet, respectively.  The coefficients $\eta^{-4/\beta_0}$ and
$\eta^{-8/\beta_0}$ are due to the running from the sfermion mass
scale $m_{\tilde{f}}$ (100 GeV assumed) down to the $m_b$ scale.
Since it is always assumed in phenomenology numerical display that
only one sfermion contributes at one time, we neglect the mixing
between the operators when we use the renormalization group
equation to run $\mathcal{H}^{RPV}_{eff}$ down to the low scale.

The RPV amplitude for the decays can be written as
\begin{eqnarray}
  \mathcal{A}^{RPV}(B\to M_1M_2)&=&
  \left< M_1M_2|{\cal H}^{RPV}_{eff}|B\right>.
\end{eqnarray}
Generally, the product RPV couplings can  be complex and their
phases may induce new contribution to CP violation, which we write
as
\begin{eqnarray}
\Lambda_{ijk}\Lambda^*_{lmn} = |\Lambda_{ijk}\Lambda_{lmn}|~e^{i
\phi_{RPV}},~~~~~\Lambda^*_{ijk}\Lambda_{lmn} =
|\Lambda_{ijk}\Lambda_{lmn}|~e^{-i \phi_{RPV}}.
\end{eqnarray}
The RPV coupling constant $\Lambda \in \{\lambda, \lambda',
\lambda''\}$, and $\phi_{RPV}$ is the RPV weak phase, which may
take any value between $-\pi$ and $\pi$.

For simplicity we only consider the vertex corrections and the hard
spectator scattering in the RPV decay amplitudes.  We ignore the RPV
penguin contributions,  which are expected to be small even compared to the SM penguin
 amplitudes,  this follows from the smallness of the relevant RPV
couplings compared to the SM gauge couplings. Thus, the bounds on
the RPV couplings are insensitive to the inclusion of the RPV
penguins
  \cite{Bhattacharyya:2002uv}. We also neglected the annihilation contributions in
the RPV amplitudes. After Fierz  transformations,  the  relevant
NP operators due to  squark exchanges are
$(\bar{u}\gamma_\mu(1+\gamma_5)q)(\bar{b}\gamma^\mu(1+\gamma_5)u)$
and
$(\bar{u}\gamma_\mu(1-\gamma_5)u)(\bar{b}\gamma^\mu(1+\gamma_5)q)$.
The factorized matrix element of these new RPV operators  is given
as follows
\begin{eqnarray}
A'_{M_1M_2}&\equiv& \langle M_2|(\bar{u}\gamma_\mu(1+\gamma_5)q)
 |0\rangle
 \langle M_1|(\bar{b}\gamma^\mu(1+\gamma_5)u)|B \rangle,\\
&=&-i\left \{\begin{array}{ll}
 m_B^2f_{M_2}F_0^{B\to M_1}(0), &\mbox{if $M_1=M_2=P$},
 \\
 m^2_Bf_{M_2}A_0^{B\to M_1}(0), &\mbox{if
$M_1=V,~M_2=P$},
\\
m^2_Bf_{M_2}F_+^{B\to M_1}(0), &\mbox{if $M_1=P,~M_2=V$},
 \\
m^2_Bf_{M_2}A_0^{B\to M_1}(0), &\mbox{if $M_1=M_2=V$ and $h=0$},
 \\
m_Bm_{M_2}f_{M_2}F^{B\to V_1}_{\mp}(0),  &\mbox{if $M_1=M_2=V$ and
$h=\pm$},
 \end{array}
 \right.
 \end{eqnarray}
with
\begin{eqnarray}
F^{B\to V_1}_{\pm}(q^2)\equiv\left(1+\frac{m_{V_1}}{m_B}\right)A_1^{B\to
V_1}(q^2)\mp\left(1-\frac{m_{V_1}}{m_B}\right)V^{B\to V_1}(q^2).
 \end{eqnarray}
Using QCDF,  we can obtain the RPV amplitudes of $B_s\to
K^{(*)-}K^{(*)+}$, $K^{(*)-}\pi^{+}$, $K^{(*)-}\rho^{+}$ decays.
There are two independent RPV amplitudes,  given by
\begin{eqnarray}
\mathcal{A}^{RPV}(\bar{B}_s \rightarrow K^{+}
K^-)&=&-\frac{\lambda''^*_{131}\lambda''_{121}}
{8m^2_{\tilde{d}}}\eta^{-4/\beta_0}F_{K \overline{K}}A'_{K
\overline{K}}-
\frac{\lambda'^*_{i13}\lambda'_{i12}}{8m^2_{\tilde{e}_{Li}}}
\eta^{-8/\beta_0}r^{K^+}_\chi N'(K) A'_{K \overline{K}},\label{AKKRPV}\\
\mathcal{A}^{RPV}(\bar{B}_s \rightarrow K^{+}
\pi^-)&=&-\frac{\lambda''^*_{132}\lambda''_{112}}
{8m^2_{\tilde{s}}}\eta^{-4/\beta_0}F_{K \pi}A'_{K \pi}-
\frac{\lambda'^*_{i13}\lambda'_{i11}}{8m^2_{\tilde{e}_{Li}}}\eta^{-8/\beta_0}
r^{\pi^+}_\chi N'(\pi)  A'_{K \pi}.\label{AKpiRPV}
\end{eqnarray}
with $N'(M_2)=1$ if $M_2=P$  and $N'(M_2)=0$ if $M_2=V$.
$F_{M_1M_2}$ is defined as
\begin{eqnarray}
  F_{M_1M_2}&\equiv& 1-\frac{1}{N_c}+\frac{\alpha_s}{4\pi}
  \frac{C_F}{N_c}\Big[V'(M_2)+\frac{4\pi^2}{N_c}H'(M_1M_2)\Big],
\end{eqnarray}
where  $V'(M_2)$ and $H'(M_1M_2)$ are the one-loop vertex
corrections and  hard spectator interactions for the new RPV operators, respectively.
The RPV amplitudes for $\overline{B}_s\to K^{+}K^{*-}$,
$\overline{B}_s\to K^{*+}K^{-}$ and  $\overline{B}_s\to
K^{*+}K^{*-}$ are obtained from Eq. (\ref{AKKRPV}) by replacing
$(K\overline{K})\to(K\overline{K}^*)$,
$(K\overline{K})\to(K^*\overline{K})$ and
$(K\overline{K})\to(K^*\overline{K}^*)$, respectively. The RPV
amplitudes for $\overline{B}_s\to K^{+}\rho^{*-}$,
$\overline{B}_s\to K^{*+}\pi^{-}$ and  $\overline{B}_s\to
K^{*+}\rho^{*-}$ are obtained from Eq. (\ref{AKpiRPV}) by replacing
$(K\pi)\to(K\rho^*)$, $(K\pi)\to(K^*\pi)$ and $(K\pi)\to(K^*\rho)$,
respectively.

As for $V'(M_2)$ and $H'(M_1M_2)$ for $M_1M_2= PP,PV,VP,VV$ cases,
the explicit results are same as these of SM operator
 $(\bar{u}\gamma_\mu(1-\gamma_5)q)(\bar{b}\gamma^\mu(1-\gamma_5)u)$ except
 ones for $B\to VV$ and $h=\pm$ case. And we get
\begin{eqnarray}
&&V'^-(V)=0,~~~~  V'^+(V)=V^-(V).\\
 && H'^-(VV)=0,~~~~ H'^+(VV)=-H^-(VV).
\end{eqnarray}

\subsection{The total decay amplitude}
From the above discussions,  the total decay amplitude are then
given as
\begin{eqnarray}
\mathcal{A}(B_{s}\to M_1 M_2)=\mathcal{A}^{SM}(B_{s}\to M_{1}
M_{2})+\mathcal{A}^{RPV}(B_{s}\to M_{1} M_{2}). \label{amp}
\end{eqnarray}
The corresponding branching ratios read
\begin{eqnarray}
\mathcal{B}(B_{s}\to M_1 M_2)=\frac{\tau_{B_s} |p_c |}{8\pi
m^2_{B_s}}\left|\mathcal{A}(B_s\rightarrow M_1 M_2)\right|^2,
\end{eqnarray}
where $\tau_{B_{s}}$ is the $B_{s}$ lifetime, $|p_c|$ is the center
of mass momentum in the center of mass frame of the $B_s$ meson. In
the $B\to VV$ decay, the two vector mesons have the same helicity,
therefore three different polarization states are possible, one
longitudinal and two transverse, and we define the corresponding
amplitudes as $\mathcal{A}_{0,\pm}$.  Transverse
$(\mathcal{A}_{\parallel,\perp})$ and helicity $(\mathcal{A}_{\pm})$
amplitudes are related by
$\mathcal{A}_{\parallel,\perp}=\frac{\mathcal{A}_+\pm\mathcal{A}_-}{\sqrt{\mathcal{A}}}$.
Then we have
\begin{eqnarray}
\left|\mathcal{A}(B\to
VV)\right|^2&=&|\mathcal{A}_0|^2+|\mathcal{A}_+|^2+|\mathcal{A}_-|^2
=|\mathcal{A}_0|^2+|\mathcal{A}_\parallel|^2+|\mathcal{A}_\perp|^2.
\end{eqnarray}

The longitudinal polarization fraction $f_L$ is defined by
\begin{eqnarray}
f_{L}(B\to VV)&=&\frac{\Gamma_L}{\Gamma}=\frac{|\mathcal{A}_0|^2}
{|\mathcal{A}_0|^2+|\mathcal{A}_\parallel|^2+|\mathcal{A}_\perp|^2}.
\end{eqnarray}

For the CPA of neutral $B$ meson decays, there is an additional
complication due to $B^0-\bar{B}^0$ mixing. There are four cases
that one encounters for neutral $B^0$ decays, as discussed in Ref.
 \cite{Gronau:1989zb,Soto:1988hf,Palmer:1994ec,Ali:1998gb}.
\begin{itemize}
\item \textbf{Case (i)}:
$B^0\to f, \bar{B}^0\to \bar{f}$, where $f$ or $\bar{f}$ is not a
common final state of $B^0$ and $\bar{B}^0$, for example $B^0_s\to
K^-\pi^+,K^-\rho^+,K^{*-}\pi^+,K^{*-}\rho^+$.
\item \textbf{Case (ii)}:
$B^0\to (f=\bar{f})\leftarrow\bar{B}^0$ with $f^{CP}=\pm f$,
involving final states which are CP eigenstates, i.e., decays such
as $B^0_s\to K^-K^+$.
\item \textbf{Case (iii)}:
$B^0\to (f=\bar{f})\leftarrow\bar{B}^0$ with $f^{CP}\neq\pm f$,
involving final states which are not CP eigenstates. They include
decays such as $B^0\to (VV)^0$, as the $VV$ states are not  CP
eigenstates.
\item \textbf{Case (iv)}:
 $B^0\to (f\&\bar{f})\leftarrow \bar{B}^0$ with $f^{CP}\neq
f$, i.e., both $f$ and $\bar{f}$ are common final states of $B^0$
and $\bar{B}^0$, but they are not CP eigenstates. Decays
$B^0_s(\bar{B}^0_s)\to K^{*-}K^+,K^{-}K^{*+}$  belong to this case.
\end{itemize}

For CP case (i) decays, there is only  direct CPA
$\mathcal{A}_{CP}^{dir}$ since no mixing is  involved for these
decays. For cases (ii) and (iii), their CPA would involve
$B^0-\bar{B}^0$ mixing. The direct CPA $\mathcal{A}_{CP}^{dir}$ and
the mixing-induced CPA $\mathcal{A}_{CP}^{mix}$ are defined
as\footnote{ We use a similar sign convention to that of
\cite{Fleischer:2005vz} for self-tagging $B^0$ and charged $B$
decays.}
\begin{eqnarray}
\mathcal{A}_{CP}^{k,dir}(B^0\to
f)=\frac{\left|\lambda_k\right|^2-1}{\left|\lambda_k\right|^2+1},~~
\mathcal{A}_{CP}^{k,mix}(B^0\to
f)=\frac{2\mbox{Im}(\lambda_k)}{\left|\lambda_k\right|^2+1},
\end{eqnarray}
where  $k=0,\parallel,\perp$ for $B\to VV$ decays and $k=0$ for
$B\to PP,PV$ decays, in addition,
$\lambda_k=\frac{q}{p}\frac{\mathcal{A}_k(\overline{B}^0\rightarrow
\bar{f})}{\mathcal{A}_k(B^0\rightarrow f)}$ for CP case (i) and
$\lambda_k=\frac{q}{p}\frac{\mathcal{A}_k(\overline{B}^0\rightarrow
f)}{\mathcal{A}_k(B^0\rightarrow f)}$ for CP cases (ii) and (iii).

Case (iv) also involves mixing but requires additional formulas.
Here one needs the four time-dependent decay widths for $B^0(t)\to
f$, $\bar{B}^0(t)\to \bar{f}$, $B^0(t)\to \bar{f}$ and
$\bar{B}^0(t)\to f$
 \cite{Gronau:1989zb,Soto:1988hf,Palmer:1994ec,Ali:1998gb}. These
time-dependent widths can be expressed by four basic matrix elements
 \cite{Palmer:1994ec}
\begin{eqnarray} g&=&\langle
f|\mathcal{H}_{eff}|B^0\rangle,~~~~h=\langle
f|\mathcal{H}_{eff}|\bar{B}^0\rangle, \nonumber\\
\bar{g}&=&\langle
\bar{f}|\mathcal{H}_{eff}|\bar{B}^0\rangle,~~~\bar{h}=\langle
\bar{f}|\mathcal{H}_{eff}|B^0\rangle,
\end{eqnarray}
which determine the decay matrix elements of $B^0\to f\&\bar{f}$ and
 $\bar{B}^0\to f\&\bar{f}$ at $t=0$. We will also study the
following quantities
\begin{eqnarray}
&&\mathcal{A}_{CP}^{k,dir}(B^0\&\bar{B}^0\to
f)=\frac{\left|\lambda'_k\right|^2-1}{\left|\lambda'_k\right|^2+1},~~
\mathcal{A}_{CP}^{k,mix}(B^0\&\bar{B}^0\to
f)=\frac{2\mbox{Im}(\lambda'_k)}{\left|\lambda'_k\right|^2+1},\\
&&\mathcal{A}_{CP}^{k,dir}(B^0\&\bar{B}^0\to
\bar{f})=\frac{\left|\lambda''_k\right|^2-1}{\left|\lambda''_k\right|^2+1},~~
\mathcal{A}_{CP}^{k,mix}(B^0\&\bar{B}^0\to
\bar{f})=\frac{2\mbox{Im}(\lambda''_k)}{\left|\lambda''_k\right|^2+1},
\end{eqnarray}
with $\lambda'_k=\frac{q}{p}(h/g)$ and
$\lambda''_k=\frac{q}{p}(\bar{g}/\bar{h})$. The signature of CP
violation is $\Gamma(\bar{B}^0(t)\to \bar{f}) \neq \Gamma(B^0(t)\to
f)$ and $\Gamma(\bar{B}^0(t)\to f) \neq \Gamma(B^0(t)\to \bar{f})$,
which means that $\mathcal{A}_{CP}^{k,dir}(B^0\&\bar{B}^0\to f)$
$\neq$ $-\mathcal{A}_{CP}^{k,dir}(B^0\&\bar{B}^0\to \bar{f})$ and/or
$\mathcal{A}_{CP}^{k,mix}(B^0\&\bar{B}^0\to f)$ $\neq$
$-\mathcal{A}_{CP}^{k,mix}(B^0\&\bar{B}^0\to \bar{f})$.

\subsection{Input Parameters}

The input parameters are collected in Table I. In our numerical
results, we will use the input parameters which are varied
randomly within $1\sigma$ range.
\begin{table}[b]
\centerline{\parbox{17cm}{\small Table I: Default values of the
input parameters and the $\pm1 \sigma$ error ranges for the
sensitive parameters used in our numerical calculations.}}
\vspace{0.3cm}
\begin{center}
\begin{tabular}{lr}\hline\hline
$m_{_{B_s}}=5.366~{\rm
GeV},~m_{_{K^{*\pm}}}=0.892~{\rm GeV},~m_{_{K^\pm}}=0.494~{\rm GeV},$\\
$m_{_{\pi^\pm}}=0.140{\rm GeV},~m_{\rho}=0.775~{\rm
GeV},~\overline{m}_b(\overline{m}_b)=(4.20\pm0.07)~{\rm GeV},$\\
$\overline{m}_u(2{\rm GeV})=(0.0015\sim 0.003)~{\rm
GeV},~\overline{m}_d(2{\rm GeV})=(0.003\sim 0.007)~{\rm GeV},$ &\\
$\overline{m}_s(2{\rm GeV})=(0.095\pm0.025)~{\rm
GeV},~\tau_{_{B_d}}=(1.530\pm0.009)~ps,~\tau_{_{B_s}}=(1.437^{+0.030}_{-0.031})~ps.$&
\cite{PDG}\\\hline
$|V_{ud}|=0.97430\pm0.00019,~|V_{us}|=0.22521^{+0.00083}_{-0.00082},~|V_{ub}|=0.00344^{+0.00022}_{-0.00017},$&\\
$|V_{cd}|=0.22508^{+0.00084}_{-0.00082},~|V_{cs}|=0.97350^{+0.00021}_{-0.00022},~|V_{cb}|=0.04045^{+0.00106}_{-0.00078},$&\\
$|V_{td}|=0.00841^{+0.00035}_{-0.00092},~|V_{ts}|=0.03972^{+0.00115}_{-0.00077},~|V_{tb}|=0.999176^{+0.000031}_{-0.000044},$&\\
$\alpha=\left(90.7^{+4.5}_{-2.9}\right)^\circ,~$$\beta=\left(21.7^{+1.0}_{-0.9}\right)^\circ,~$
$\gamma=\left(67.6^{+2.8}_{-4.5}\right)^\circ.$&
\cite{CKMfit}\\\hline
$f_{K}=0.160~{\rm GeV},~f_{K^*}=(0.217\pm0.005)~{\rm GeV},~f^{\perp}_{K^*}=(0.156\pm0.010)~{\rm GeV},$\\
$f_{\pi}=0.131~{\rm GeV},~f_{\rho}=(0.205\pm0.009)~{\rm GeV},~f^{\perp}_{\rho}=(0.147\pm0.010)~{\rm GeV},$&\\
$A^{B_{s}\to  K^*}_{0}(0)=0.360\pm0.034,~A_1^{B_{s}\rightarrow
K^{\ast}}(0)=0.233\pm 0.022,
~A_2^{B_{s}\rightarrow K^{\ast}}(0)=0.181\pm 0.025,$\nonumber\\
 $ V^{B_{s}\rightarrow K^{\ast}} (0) =0.311\pm 0.026,~F^{B_{s}\to K}_{0}(0)=0.30^{+0.04}_{-0.03}.$
& \cite{BallZwicky,Melic}\\\hline
$f_{B_{s}}=(0.245\pm0.025)~{\rm GeV}.$
& \cite{Lubicz:2008am}\\\hline
$\lambda_B=(0.46\pm0.11)$ GeV. &  \cite{Braun:2003wx}\\\hline
$\alpha^\pi_1=0,~\alpha^\pi_2=0.20\pm0.15,~\alpha^\rho_1=0,~\alpha^\rho_2=0.1\pm0.2,$\\
$\alpha^K_1=0.2\pm0.2,~\alpha^K_2=0.1\pm0.3,~\alpha^{K^*}_1=0.06\pm0.06,~\alpha^{K^*}_2=0.1\pm0.2.$
&  \cite{Beneke:2003zv,Beneke:2006hg} \\\hline\hline
\end{tabular}
\end{center}
\end{table}
The Wilson coefficients $C_i$ are evaluated at scales $\mu=m_b$
\cite{coeff}. For hard spectator scattering, we take
$\mu_h=\sqrt{\Lambda_{QCD}m_b}$.
 When we study the RPV effects, we  consider
only one RPV coupling product to contribute at one time, neglecting
the interferences between different RPV coupling products, but
keeping their interferences  with the SM amplitude. We assume that
the masses of the sfermions are 100 GeV. For other values of the
sfermion masses, the bounds on the couplings derived in this paper
can be easily obtained by scaling them by factor
$\tilde{f}^2\equiv(\frac{m_{\tilde{f}}}{100~\rm{GeV}})^2$.

\section{Numerical results and Analysis}

Now we are ready to present our numerical results and analysis.
First, we will show our estimations in the SM with the parameters listed in Table I
 and compare with the relevant experimental data. Then, we will consider the RPV
effects and  constrain the relevant RPV couplings from  the
experimental data. Using the constrained parameter spaces, we will
give the RPV SUSY predictions for the branching ratios, the CP
asymmetries
 and the longitudinal polarization fractions,
which have not been measured yet in $B_s\to K^{(*)-}K^{(*)+}$,
$K^{(*)-}\pi^{+}$, $K^{(*)-}\rho^{+}$ systems.

For CP case (i) decays $B_s\to K^{(*)-}\pi^+$ and $
K^{(*)-}\rho^+$, we will study the CP averaged branching ratios
($\mathcal{B}$), $\mathcal{A}^{dir}_{CP}$ and the longitudinal
polarization fractions ($f_L$). For CP cases (ii), (iii) and (iv)
decays $B_s\rightarrow K^{(*)-}K^{(*)+}$, we  will also study
$\mathcal{A}^{mix}_{CP}$
 besides $\mathcal{B}$,
$\mathcal{A}^{dir}_{CP}$ and $f_L$. For CPA of $B_s\to K^{*-}K^{*+},
K^{*-}\rho^+$, we only study the longitudinal direct
 CPA ($\mathcal{A}^{L,dir}_{CP}$) and longitudinal
mixing-induced CPA ($\mathcal{A}^{L,mix}_{CP}$). The numerical
results in the SM are presented in Table II. The detailed error
estimates corresponding to the different types of theoretical
uncertainties have been already studied  in Refs.
 \cite{Beneke:2003zv,Beneke:2006hg}, and our SM results of
$\mathcal{B}$, $\mathcal{A}^{dir}_{CP}$ and $f_L$ are consistent
with the ones in Refs.  \cite{Beneke:2003zv,Beneke:2006hg}.

\begin{table}[ht]
\centerline{\parbox{16cm}{{Table II: The SM predictions   for
$\mathcal{B}$ (in units of $10^{-5}$), $\mathcal{A}^{dir}_{CP}$,
and $\mathcal{A}^{mix}_{CP}$ in $B_s\to K^-K^+,
K^{*-}K^+,K^-K^{*+} $, $K^-\pi^{+}$, $K^{*-}\pi^{+},K^-\rho^+$
decays within QCDF. $B_s\&\overline{B}_s\rightarrow K^{*-}K^{+}$
denotes that $B^0_s$ and $\overline{B}^0_s$ decay to the same
final state $K^{*-}K^{+}$. }}} \vspace{0.4cm}
\begin{center}
\begin{tabular}
{l|c|c|c}\hline\hline
 Decay modes& $\mathcal{B}$&
 $\mathcal{A}^{dir}_{CP}$&$\mathcal{A}^{mix}_{CP}$\\\hline
 $B_s\rightarrow K^-K^+$ &[0.89,~4.45]&[0.02,~0.06]&[0.21,~0.43]\\
$B_s\rightarrow K^{*-}K^+$&[0.22,~2.14]&[-0.07,~0.02]\\
$B_s\rightarrow K^-K^{*+}$&[0.21,~0.65]&[0.03,~0.10]\\
$B_s\&\overline{B}_s\rightarrow K^{*-}K^{+}$&&[-0.72,~0.34]&[-0.30,~0.04]\\
$B_s\&\overline{B}_s\rightarrow
K^-K^{*+}$&&[-0.31,~0.73]&[-0.34,~0.12]\\\hline
$B_s\rightarrow K^-\pi^+$&[0.61,~1.47]&[-0.09,~-0.05] \\
$B_s\rightarrow K^{*-}\pi^+$&[0.84,~1.72]&[0.00,~0.02]\\
$B_s\rightarrow K^-\rho^+$&[1.33,~3.51]&[-0.02,~-0.01]\\
\hline
\end{tabular}
\end{center}
\end{table}

\begin{table}[ht]
\centerline{\parbox{16cm}{{Table III: The SM predictions   for
$\mathcal{B}$ (in units of $10^{-5}$), $\mathcal{A}^{L,dir}_{CP}$,
$\mathcal{A}^{L,mix}_{CP}$ and $f_L$ in $B_s\to K^{*-}K^{*+}$,
$K^{*-}\rho^{+}$ decays within QCDF. }}} \vspace{0.4cm}
\begin{center}
\begin{tabular}
{l|c|c|c|c}\hline\hline
 Decay modes& $\mathcal{B}$&
 $\mathcal{A}^{L,dir}_{CP}$&$\mathcal{A}^{L,mix}_{CP}$&$f_L$\\\hline

$B_s\rightarrow
K^{*-}K^{*+}$&[0.39,~1.71]&$[-0.04,~0.19]$&[0.70,~0.93]&[0.38,~0.89]\\\hline

$B_s\rightarrow K^{*-}\rho^+$&[1.03,~6.23]&[-0.06,~-0.03]&&[0.86,~0.97]\\
\hline
\end{tabular}
\end{center}
\end{table}

\begin{itemize}
\item Our results of $B\to PP$ and $PV$ are obtained excluding the
uncertainties of power corrections parameterized by the quantities
$X_A$ and $X_H$. In the QCDF, the  endpoint divergent integrals
appear in the hard-scattering contributions and in the weak
annihilation contributions, which are treated with model-dependent
parameters \cite{Beneke:2003zv} $X_H\equiv(1+\varrho_H e^{i
\varphi_H})\mbox{ln}\frac{m_B}{\Lambda_h}$ and
$X_A\equiv(1+\varrho_A e^{i
\varphi_A})\mbox{ln}\frac{m_B}{\Lambda_h}$, respectively.  The
different $X_A$ values are allowed
 for the four cases $PP$, $PV$, $VP$ and $VV$.
Our results of $B\to PP,PV$ are obtained without  the
uncertainties of power corrections and set
$\varrho_A=\varrho_H=0$.
For two vector final-state meson decays,  in order to be
consistent with the longitudinal polarization fractions around
$0.5$ in the penguin-dominated decays $B\to \phi K^{*0}$ and
$\rho^+ K^{*0}$, maximal annihilation contribution are considered
($\varrho_A=0.6$ and $\varphi_A=-40^\circ$) in Ref.
 \cite{Beneke:2006hg}. We also consider the large annihilation
contribution and suggest $\varrho_H=0$, $\varrho_A=0.6\pm0.2$ and
$\varphi_A=(-40\pm10)^\circ$.   The  annihilation topology obviously
contributes to $\mathcal{B}$, $\mathcal{A}^{L,dir}_{CP}$ and
$\mathcal{A}^{L,mix}_{CP}$ besides $f_L$ in $B_s\rightarrow
K^{*-}K^{*+}$ decay. For example,
$\mathcal{A}^{L,dir}_{CP}(B^0_s\rightarrow K^{*-}K^{*+})$ receives
much larger annihilation contribution than
$\mathcal{A}^{dir}_{CP}(B^0_s\rightarrow K^-K^+)$ does.  It is also  noted that
annihilation contribution could cancel voluminous penguin
contribution in $\mathcal{A}^{L,dir}_{CP}(B^0_s\rightarrow
K^{*-}K^{*+})$.

\item For CP case (iv) $B_s\to K^{*-}K^+$ decay,  the final state can
 come both from a pure $B_s$ and a
pure $\bar{B}_s$, the amplitudes for the direct decay $B_s\to
K^{*-}K^+$ and the mixing-induced sequence $B_s\to \bar{B}_s\to
K^{*-}K^+$. We obtain $\mathcal{A}^{dir}_{CP}(B_s\&\bar{B}_s\to
K^{*-}K^+ )$ $\approx$ $-\mathcal{A}^{dir}_{CP}(B_s\&\bar{B}_s\to
K^{-}K^{*+} )$, however, $\mathcal{A}^{mix}_{CP}(B_s\&\bar{B}_s\to
K^{*-}K^+ )$ $\neq$ $-\mathcal{A}^{mix}_{CP}(B_s\&\bar{B}_s\to
K^{-}K^{*+} )$, which imply that its direct CP violation is very
small, nevertheless its CP violating effect can appear through the
interference of the direct decay $B_s\to K^{*-}K^+$ and the
mixing-induced decay $B_s\to \bar{B}_s\to K^{*-}K^+$. In addition,
the theoretical predictions for above CP asymmetry parameters
suffer large uncertainties, which are dominated by the
uncertainties of mass and the Gegenbauer moments in the expansion
of the light-cone distribution amplitudes, and also due to the
uncertainties of the form factors and the CKM matrix elements.

\item In  penguin-dominated decay $B^0_s \to K^{*+}K^{*-}$,
as transverse and longitudinal contributions can be
be of  the similar magnitude, the CP asymmetry and the polarization
fractions predictions will suffer large uncertainties. For example,
compared to $\mathcal{A}^{dir}_{CP}(B^0_s\rightarrow K^{-}K^{+})$,
$\mathcal{A}^{L,dir}_{CP}(B^0_s \to K^{*-}K^{*+})$ suffers
quite large uncertainties, which mostly come from the uncertainties
of the relevant form factors and  the weak annihilation parameter $X_A$.
$f_{L}(B^0_s \to K^{*-}K^{*+})$ has quite large allowed range
for the same reason as $\mathcal{A}^{L,dir}_{CP}(B^0_s\rightarrow
K^{*-}K^{*+})$.

\item $\mathcal{A}^{L,mix}_{CP}(B^0_s\rightarrow K^{*-}K^{*+})$  is
much larger than $\mathcal{A}^{mix}_{CP}(B^0_s\rightarrow K^-K^+)$
 in the SM. Large difference between them arises  from chirally-enhanced terms,
 which give large contribution to penguin-dominated decay modes with pseudoscalar final-states.

\item For the color-allowed tree-dominated decays $B_s\to
K^-\pi^+$, $K^{*-}\pi^+$, $K^-\rho^+$, and $K^{*-}\rho^+$, power
corrections have limited impact, and the main sources of
theoretical uncertainties in the branching ratio are CKM matrix
elements and form factors. Their $A^{dir}_{CP}$ and
$A^{L,dir}_{CP}$ can be predicted quite precisely, and found to be
very small ($\sim 10^{-2}$) due to small penguin amplitudes. The
uncertainty of $f_L(B_s\to K^{*-}\rho^+)$ is mostly due to the
uncertainties of form factors.

\end{itemize}

Now we turn to the RPV effects in $B_s\to K^{(*)-}K^{(*)+}$,
$K^{(*)-}\pi^{+}$, and  $K^{(*)-}\rho^{+}$ decays. There are two RPV
coupling products, $\lambda''^*_{131}\lambda''_{121}$ and
$\lambda'^*_{i13}\lambda'_{i12}$ contributing to four $B_s\to
K^{(*)-}K^{(*)+}$ modes, which involve the quark level process $b\to
u\bar{u}s$. Four decays $B_s\to K^{(*)-}\pi^{+}$, $K^{(*)-}\rho^{+}$
are due to $b\to u\bar{u}d$ at the quark level, and the relevant RPV
coupling products are $\lambda''^{*}_{132}\lambda''_{112}$ and
$\lambda'^*_{i13}\lambda'_{i11}$. We use the
 experimental results shown in Eq. (\ref{data}) to
constrain the relevant RPV parameters.
\begin{figure}[ht]
\begin{center}
\includegraphics[scale=0.6]{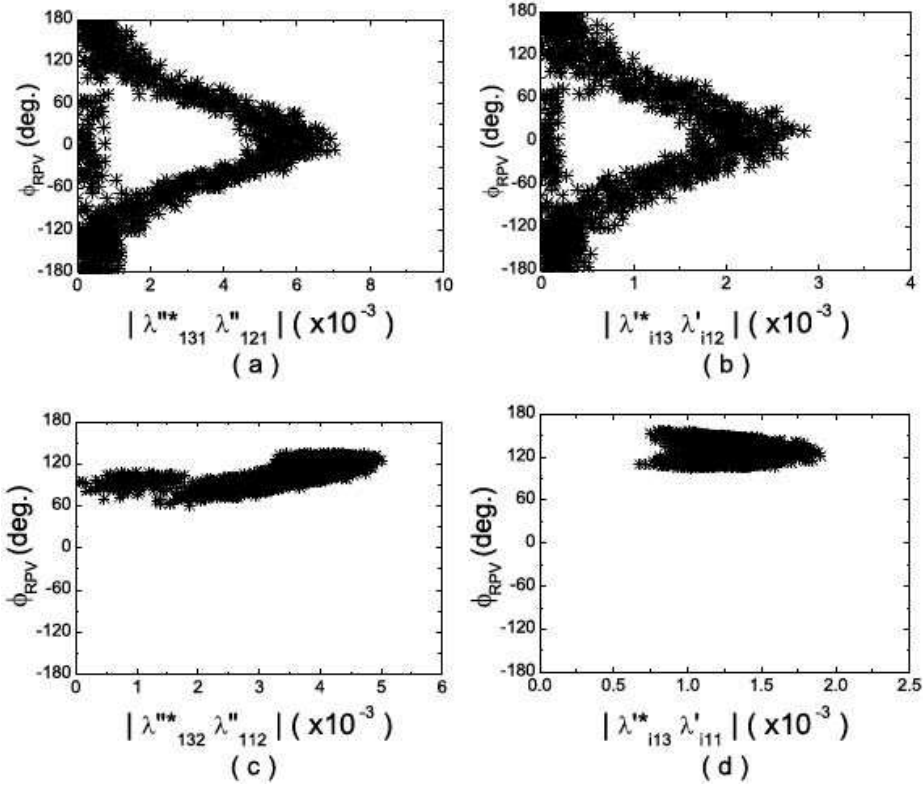}
\end{center}
\vspace{-0.6cm}
 \caption{ The allowed parameter spaces for the relevant
 RPV couplings constrained by $B_s\to K^{-}K^{+}$ and $K^{-}\pi^{+}$. $\phi_{RPV}$ denotes the RPV weak phase.}
 \label{rpvbounds}
\end{figure}

Our bounds on  $\lambda''^*_{131}\lambda''_{121}$ and
$\lambda'^*_{i13}\lambda'_{i12}$ are demonstrated in Fig.
\ref{rpvbounds} (a-b) by using the experimental measurement  of
$\mathcal{B}(B_s\to K^-K^+)$ within $1\sigma$ error-bar range. From Fig.
\ref{rpvbounds} (a-b), we find that the RPV weak phases of
$\lambda''^*_{131}\lambda''_{121}$ and
$\lambda'^*_{i13}\lambda'_{i12}$ are not much constrained, but the
modulus of the two RPV coupling products can be tightly upper
limited. Since the SM prediction ranges of $\mathcal{A}_{CP}^{dir}$
($\mathcal{B}$) in $B_s\rightarrow K^-\pi^+$ decay summarized in
Table II is a little smaller (larger) than the corresponding
 measurements within $1\sigma$  by CDF shown in Eq.
 (\ref{data}),
the allowed ranges of $\lambda''^{*}_{132}\lambda''_{112}$ and
$\lambda'^*_{i13}\lambda'_{i11}$ are strongly restricted by these
experimental data. We obtain
$|\lambda''^{*}_{132}\lambda''_{112}|\in[0.22,~4.86]\times10^{-3}$
and its phase  $\phi_{RPV}\in [80^\circ,123^\circ]$. However,  we could not
 find the allowed space of $\lambda'^*_{i13}\lambda'_{i11}$ within  $1\sigma$ error-bar
 of the experimental bounds.  Within  $2\sigma$ error-bar of the experimental data,
 one can find the allowed spaces of these two RPV coupling products which
   are given in Fig. \ref{rpvbounds} (c-d). One can find that the RPV weak phases only have the
positive values, the RPV weak phase of
$\lambda''^{*}_{132}\lambda''_{112}$ lies in
$[60^\circ,139^\circ]$ and the phase of
$\lambda'^*_{i13}\lambda'_{i11}$ lies in $[104^\circ,158^\circ]$.
Furthermore,  the strengths of the two RPV coupling products are
restricted strongly, which
 are summarized in Table IV.
 For comparison, the
existing bounds on these quadric coupling products, which obtain
from $B_{u,d}$ decays of  the same quark level process
\cite{Ghosh:2001mr,Yang:2005es}  are also listed.
\begin{table}[tbp]
\centerline{\parbox{16cm}{Table IV: Bounds on the relevant RPV
couplings  by $B_s\to K^{-}K^{+}$, $K^{-}\pi^{+}$
 decays for 100 GeV sfermions. Previous bounds
are  listed for comparison. }} \vspace{0.5cm}
\begin{center}{\scriptsize
\begin{tabular}{c|l|l|l}\hline\hline
Couplings&~~~~~~~Our bounds~~ [Process]& Bounds I~~  [Process] ~~~~
 \cite{Yang:2005es} & Bounds II~~   [Process]~~~~
 \cite{Ghosh:2001mr}\\\hline $|\lambda''^*_{131}\lambda''_{121}|
$&$\leq7.01\times 10^{-3}~[B_s\to K^+K^-]$&$^{[0.61,~4.6]\times
10^{-3}}_{[5.6,~~7.2]\times 10^{-3}}~[B\to \pi K]$ &$\leq1.54\times
10^{-2}~[B_{u}\to
K^-\pi^0]$  \\
 $|\lambda'^*_{i13}\lambda'_{i12}|$&$\leq2.84\times
10^{-3}~[B_s\to K^+K^-]$ &$[0.36,~1.1]\times 10^{-3}~[B\to \pi K]$
&$\leq2.71\times 10^{-3}~[B_u\to K^-\pi^0]$
\\\hline
$|\lambda''^{*}_{132}\lambda''_{112}| $&$\leq5.01\times
10^{-3}~[B_s\to K^+\pi^-]$&$[0.54,~~2.9]\times 10^{-3}~[B_d\to \pi
\pi]$ &$\leq4.69\times 10^{-3}~[B_d\to
\pi^+ \pi^-]$ \\
$|\lambda'^*_{i13}\lambda'_{i11}|$&$[0.67,~1.90]\times
10^{-3}~[B_s\to K^+\pi^-] $&$[0.27,~0.77]\times 10^{-3}~[B_d\to \pi
\pi]$&$\leq1.90\times 10^{-3}~[B_d\to \pi^-\pi^+]$
\\\hline\hline
\end{tabular}}
\end{center}
\end{table}
Note that, previous bounds-I of Ref.  \cite{Yang:2005es} are
obtained by considering the experimental constraints of all
relevant decay modes at the same time, so the allowed RPV coupling
spaces are very narrow. In Ref.  \cite{Ghosh:2001mr}, the bounds
are given through experimental restraints  mode by mode.
Our bounds on $\lambda''^*_{131}\lambda''_{121}$,
$\lambda'^*_{i13}\lambda'_{i12}$
 and $\lambda''^*_{132}\lambda''_{112}$
 are consistent with the existing
ones in Refs.  \cite{Ghosh:2001mr}, and  just a little weaker than
these in Ref.  \cite{Yang:2005es} which are obtained from many correlated
experimental constraints. Our bound of
$|\lambda'^*_{i13}\lambda'_{i11}|$ also  consists
 with one from Ref.  \cite{Ghosh:2001mr}, however, there is only very narrow
overlap  between range $[0.27,0.77]\times10^{-3}$ in Ref.
 \cite{Yang:2005es} and ours $[0.67,1.90]\times10^{-3}$, therefore,
it should be of order   $10^{-4}$  if
$|\lambda'^*_{i13}\lambda'_{i11}|$ can survive.

Next, we will use the constrained parameter spaces from $B_s\to
K^{-}K^{+}$ and  $K^{-}\pi^{+}$ decays, as shown in Fig.
\ref{rpvbounds}, to predict the RPV effects on the other
quantities which have not been measured yet in $B_s\to
K^{(*)-}K^{(*)+}$, $K^{(*)-}\pi^{+}$ and  $K^{(*)-}\rho^{+}$
decays. With the expressions for $\mathcal{B}$,
$\mathcal{A}^{dir}_{CP}$, $\mathcal{A}^{mix}_{CP}$ and $f_L$, we
perform a scan through the input parameters and the new
constrained RPV coupling spaces, and then the allowed ranges for
$\mathcal{B}$, $\mathcal{A}^{dir}_{CP}$, $\mathcal{A}^{mix}_{CP}$
and $f_L$ are obtained with different RPV couplings, which satisfy
relevant experimental constraints of $B_s$ decays given in Eq.
(\ref{data}).  The numerical results for $B_s \rightarrow
K^{(*)-}K^{(*)+}$ and $B_s \rightarrow K^{(*)-}\pi^+,
K^{(*)-}\rho^+$  are summarized in Table V and Table VI,
respectively.
\begin{table}[htbp]
\centerline{\parbox{16.5cm}{{Table V: The theoretical predictions
of $B_s \rightarrow K^{(*)-}K^{(*)+}$ for $\mathcal{B}$ (in units
of $10^{-5}$), $\mathcal{A}^{dir}_{CP}$, $\mathcal{A}^{mix}_{CP}$
and $f_L$  with the allowed regions of the different RPV
couplings. }}} \vspace{0.6cm}
\begin{center}\small{
\begin{tabular}
{l|l|l}\hline\hline
&$\lambda''^{*}_{131}\lambda''_{121}$&$\lambda'_{i12}\lambda'^{*}_{i13}$\\\hline
$\mathcal{B}(B_s \rightarrow K^{*-}K^+)$&$[0.33,~10.56]$&$[0.32,~12.89]$\\
$\mathcal{B}(B_s \rightarrow K^{-}K^{*+})$&$[0.01,~28.88]$&\\
$\mathcal{B}(B_s \rightarrow K^{*-}K^{*+})$
&$[0.29,~27.23]$&\\\hline
$\mathcal{A}^{dir}_{CP}(B_s\rightarrow K^-K^+)$ &$[-0.50,~0.50]$&$[-0.25,~0.23]$\\
$\mathcal{A}^{dir}_{CP}(B_s\rightarrow K^{*-}K^+)$ &$[-0.25,~0.50]$&$[-0.11,~0.22]$\\
$\mathcal{A}^{dir}_{CP}(B_s\rightarrow K^{-}K^{*+})$ &$[-0.98,~0.95]$&\\
$\mathcal{A}^{L,~dir}_{CP}(B_s \rightarrow
K^{*-}K^{*+})$&$[-0.30,~0.52]$ &\\\hline
$\mathcal{A}^{mix}_{CP}(B_s\rightarrow K^-K^+)$ &$[-0.97,~0.98]$&$[-0.99,~1.00]$\\
$\mathcal{A}^{mix}_{CP}(B_s\&\overline{B}_s \rightarrow K^{*-}K^+)$&$[-0.89,~0.97]$&$[-0.98,~0.55]$\\
$\mathcal{A}^{mix}_{CP}(B_s\&\overline{B}_s \rightarrow K^-K^{*+})$ &$[-0.89,~0.96]$&$[-0.98,~0.58]$\\
$\mathcal{A}^{L,~mix}_{CP}(B_s \rightarrow
K^{*-}K^{*+})$&$[-1.00,~1.00]$ &\\\hline
$f_L(B_s \rightarrow K^{*-}K^{*+})$&$[0.30,~0.97]$&\\
\hline\hline
\end{tabular}}
\end{center}
\end{table}
\begin{table}[htbp]
\centerline{\parbox{16.5cm}{{Table VI: The theoretical predictions
of $B_s \rightarrow K^{(*)-}\pi^+, K^{(*)-}\rho^+$ for
$\mathcal{B}$ (in units of $10^{-5}$), $\mathcal{A}^{dir}_{CP}$
and $f_L$ with  the allowed regions of the different RPV
couplings. }}} \vspace{0.6cm}
\begin{center}\small{
\begin{tabular}
{l|l|l}\hline\hline
&$\lambda''^{*}_{132}\lambda''_{112}$&$\lambda'^*_{i13}\lambda'_{i11}$\\\hline
$\mathcal{B}(B_s \rightarrow K^{*-}\pi^+)$&$[1.34,~11.59]$&$[4.49,~13.47]$\\
$\mathcal{B}(B_s \rightarrow K^-\rho^+)$&$[2.02,~23.32]$&\\
$\mathcal{B}(B_s \rightarrow K^{*-}\rho^+)$&$[0.40,~3.31]$&\\\hline
$\mathcal{A}^{dir}_{CP}(B_s \rightarrow
K^{*-}\pi^+)$&$[-0.04,~0.04]$&$[0.00,~0.02]$\\
$\mathcal{A}^{dir}_{CP}(B_s \rightarrow K^-\rho^+)$&$[-0.05,~0.01]$&\\
$\mathcal{A}^{L,~dir}_{CP}(B_s \rightarrow
K^{*-}\rho^+)$&$[-0.25,~0.73]$&\\\hline
 $f_L(B_s\rightarrow
K^{*-}\rho^+)$&$[0.42,~0.96]$&\\
\hline\hline
\end{tabular}}
\end{center}
\end{table}

Comparing the RPV SUSY predictions given in Table V and Table VI
to the SM values listed in Table II and Table III, we give some
remarks on the numerical results.
\begin{itemize}
\item All branching ratios can be greatly changed by the RPV
couplings compared  to the SM expectations.
\item The RPV effects on $\mathcal{A}^{dir}_{CP}(B_s \rightarrow
K^{*-}\pi^+)$ and $\mathcal{A}^{dir}_{CP}(B_s \rightarrow
K^{-}\rho^+)$ are found to be very small, but could be large for
 the direct CPA in other five $B_s
\rightarrow K^{*-}\rho^+$ and $K^{(*)-}K^{(*)+}$ decays.
\item The mixing-induced CPA in $B_s
\rightarrow K^{(*)-}K^{(*)+}$ system can be greatly enhanced by the RPV couplings
$\lambda''^*_{131}\lambda''_{121}$ and
$\lambda'^*_{i13}\lambda'_{i12}$.
\item The squark exchange couplings
$\lambda''^*_{131}\lambda''_{121}$ and
$\lambda''^{*}_{132}\lambda''_{112}$ could have significant impacts on
$f_L(B_s\rightarrow K^{*-}K^{*+})$ and $f_L(B_s\rightarrow
K^{*-}\rho^+)$, which could be decreased as low as  $0.30$ and 0.42,
respectively.
\end{itemize}

\begin{figure}[hb]
\begin{center}
\includegraphics[scale=0.7]{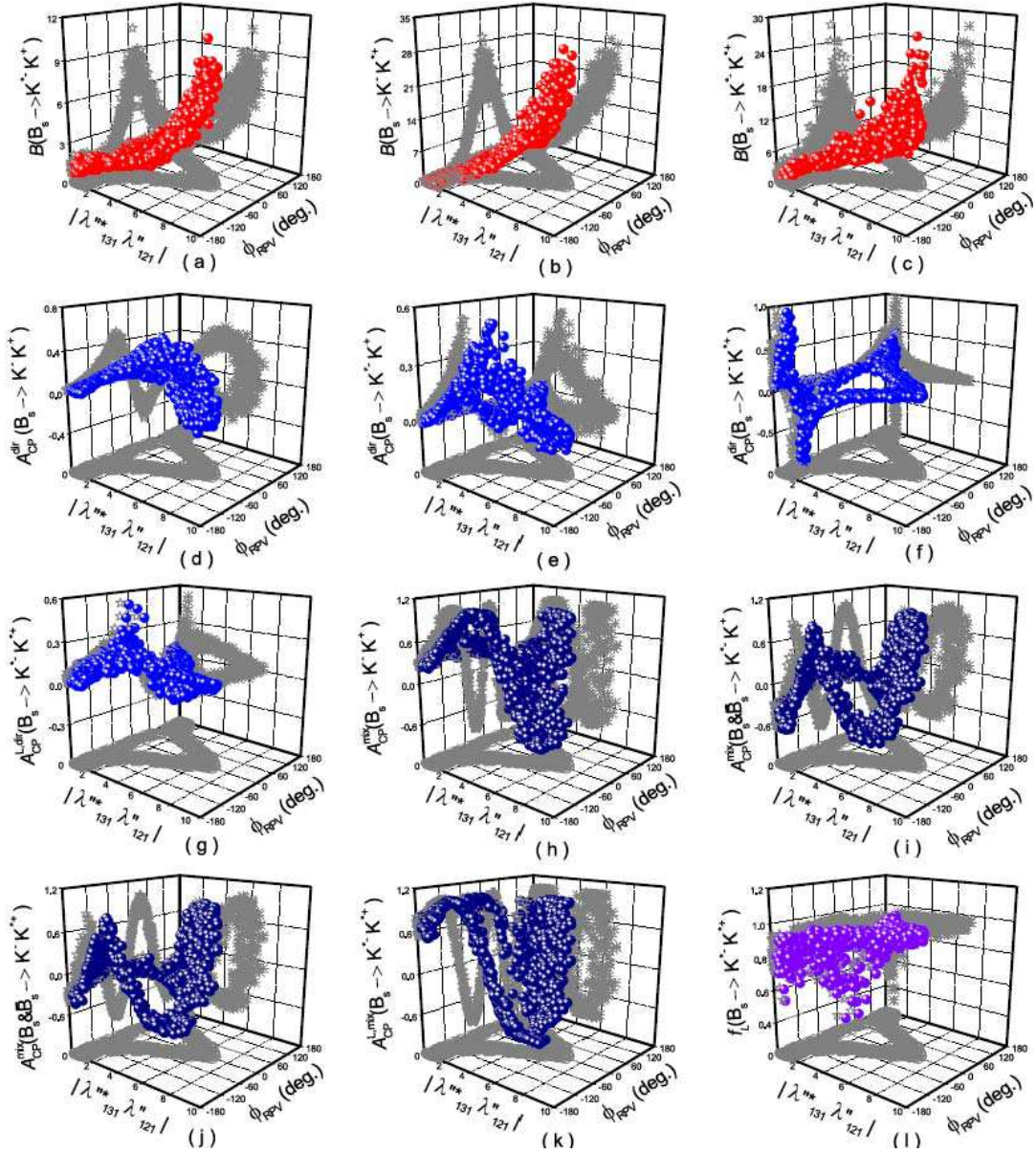}
\end{center}
\vspace{-0.6cm}
 \caption{\small The effects of RPV coupling $\lambda''^*_{131}\lambda''_{121}$
  in $B_s\to K^-K^+,K^-K^{*+},K^{*-}K^+$, $K^{*-}K^{*+}$ decays. $\mathcal{B}$  in
units of $10^{-5}$ and $|\lambda''^*_{131}\lambda''_{121}|$  in
units of $10^{-3}$.}
 \label{131121}
\end{figure}

\begin{figure}[ht]
\begin{center}
\includegraphics[scale=0.7]{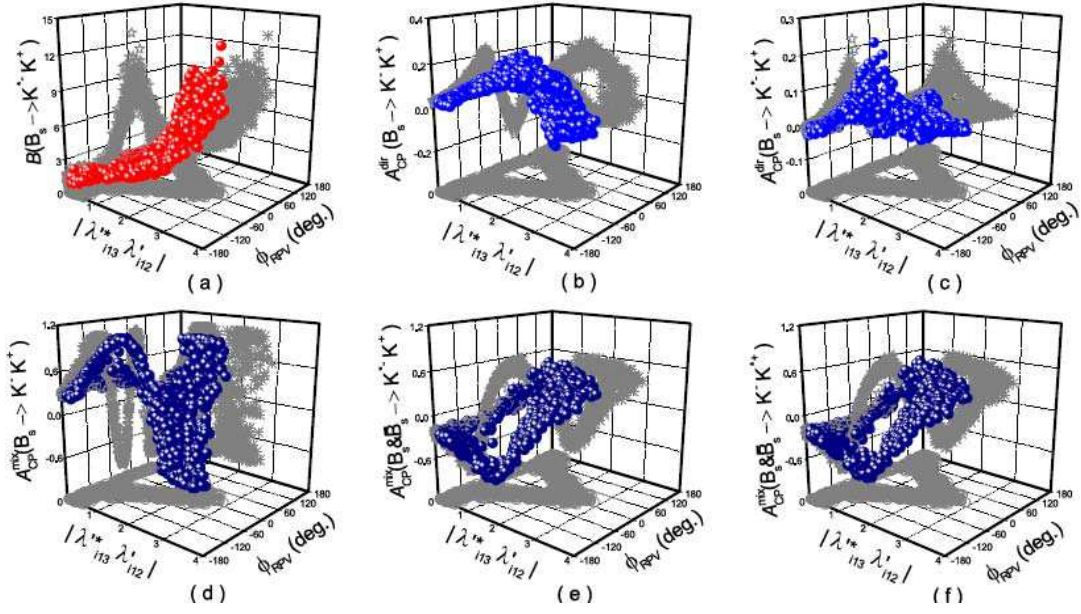}
\end{center}
\vspace{-0.6cm} \caption{\small The effects of RPV coupling $
\lambda'^*_{i13}\lambda'_{i12}$ in $B_s\to K^-K^+$, $K^-K^{*+}$
decays. $\mathcal{B}$ in units of $10^{-5}$ and $
|\lambda'^*_{i13}\lambda'_{i12}|$ in units of $10^{-3}$.}
 \label{i13i12}
\end{figure}

\begin{figure}[ht]
\begin{center}
\includegraphics[scale=0.7]{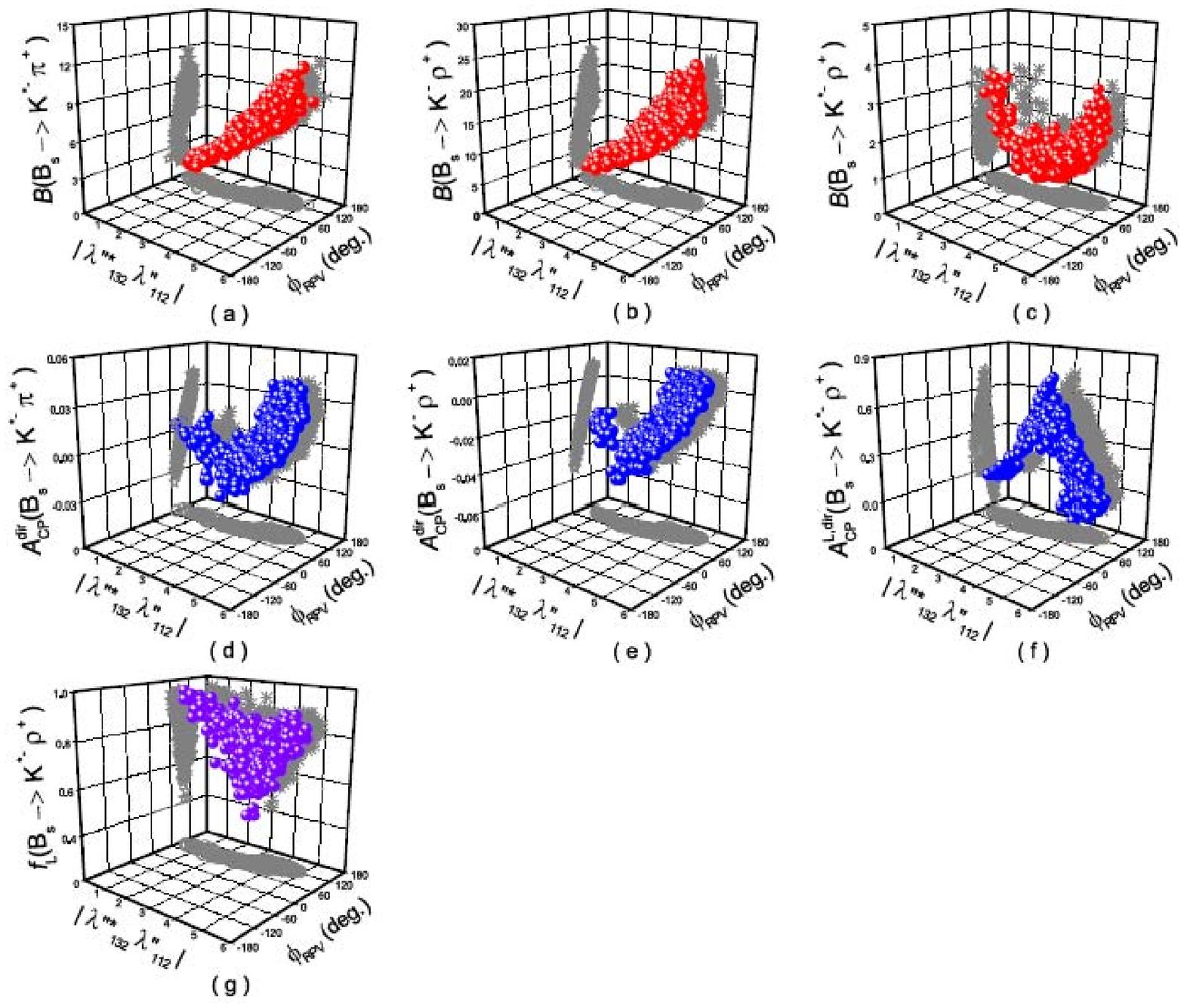}
\end{center}
\vspace{-0.6cm}
 \caption{\small The effects of RPV coupling $\lambda''^*_{132}\lambda''_{112}$ in $B_s\to K^{*-}\pi^+$, $K^-\rho^+$, $K^{*-}\rho^+$ decays.
 $\mathcal{B}$  in
units of $10^{-5}$ and $|\lambda''^*_{132}\lambda''_{112}|$ in
units of $10^{-3}$.}
 \label{112132}
\end{figure}

\begin{figure}[ht]
\begin{center}
\includegraphics[scale=0.5]{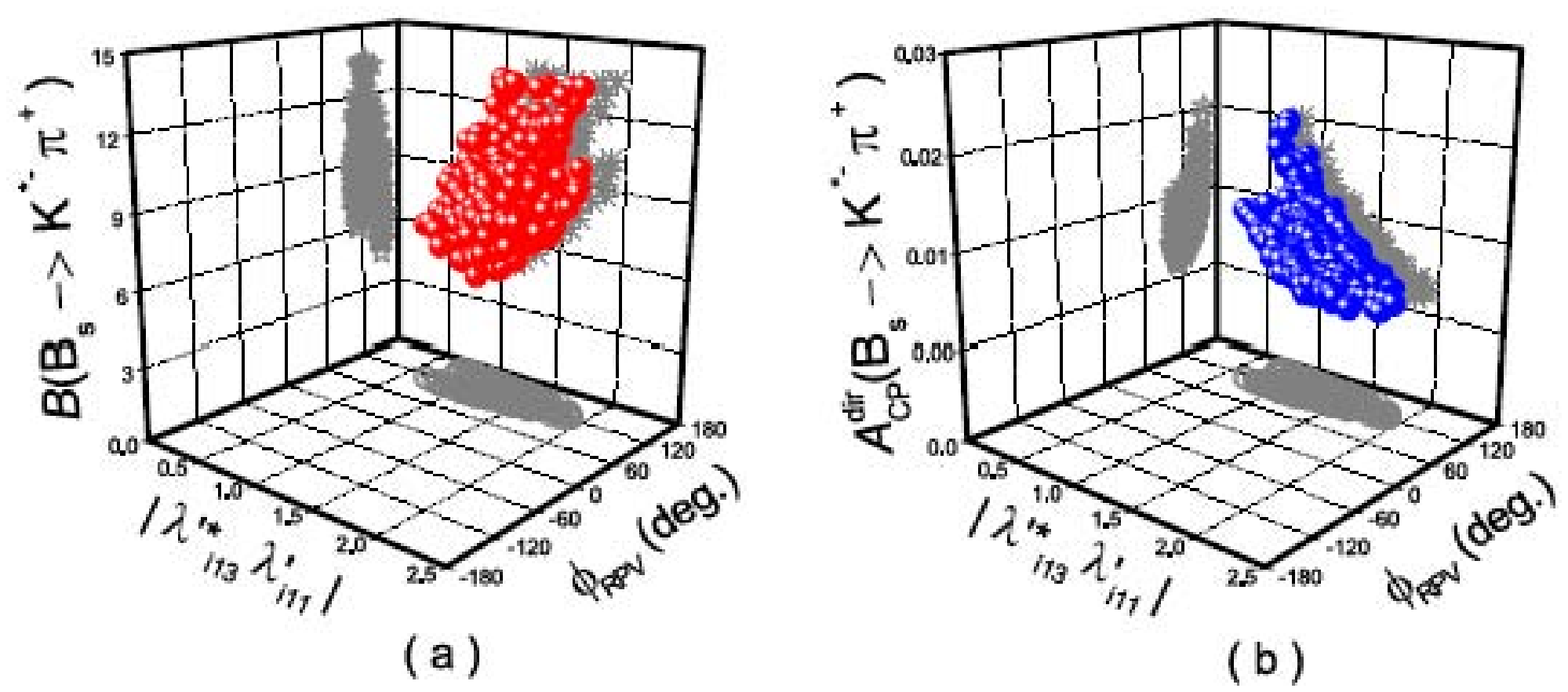}
\end{center}
\vspace{-0.6cm}
 \caption{\small The effects of RPV coupling $\lambda'^*_{i11}\lambda'_{i13}$ in $B\to K^{*-}\pi^+$ decays.
 $\mathcal{B}$ in
units of $10^{-5}$ and $|\lambda'^*_{i13}\lambda'_{i11}|$ in units
of $10^{-3}$.}
 \label{i13i11}
\end{figure}

In Figs. \ref{131121}-\ref{i13i11}, we present correlations between
the physical observable $\mathcal{B}$, $\mathcal{A}^{dir}_{CP}$,
$\mathcal{A}^{mix}_{CP}$, $f_L$ and the parameter spaces of
different RPV couplings by these  three-dimensional scatter plots.
 From Figs. \ref{131121}-\ref{i13i11}, one can see the changing trends of
the physical observables   with  the modulus and RPV weak phase
$\phi_{RPV}$. Taking  the first plot in Fig. \ref{131121}(a) as an
example, this plot shows $\mathcal{B}(B_s\rightarrow K^{*-}K^{+})$
changing  trend with RPV coupling
$\lambda''^*_{131}\lambda''_{121}$, where  projections on three
perpendicular
 planes are also given.  The $|\lambda''^*_{131}\lambda''_{121}|$-$\phi_{RPV}$
 plane displays the allowed
 regions of $\lambda''^*_{131}\lambda''_{121}$ which satisfy experimental data in
 Eq. (\ref{data}) (the same as the Fig.\ref{rpvbounds}(a)).
 The $\mathcal{B}(B_s\rightarrow K^{*-}K^{+})$-$|\lambda''^*_{131}\lambda''_{121}|$ plane shows
  that $\mathcal{B}(B_s\rightarrow K^{*-}K^{+})$ is increasing  with
$|\lambda''^*_{131}\lambda''_{121}|$, the
$\mathcal{B}(B_s\rightarrow
 K^{*-}K^{+})$-$\phi_{RPV}$ plane shows that
$\mathcal{B}(B_s\rightarrow  K^{*-}K^{+})$ is decreasing with
$|\phi_{RPV}|$.
  Additional refined measurements of $\mathcal{B}(B_s\rightarrow
 K^{-}K^{+})$ can further restrict the constrained space of
$\lambda''^*_{131}\lambda''_{121}$, thus more accurate
$\mathcal{B}(B_s\rightarrow K^{*-}K^{+})$ can be predicted.
  The following salient features
  in Figs. \ref{131121}-\ref{i13i11} are summarized as following.

\begin{itemize}

\item  Fig.\ref{131121} displays the effects of RPV coupling
$\lambda''^*_{131}\lambda''_{121}$ on $\mathcal{B}$,
$\mathcal{A}^{dir}_{CP}$, $\mathcal{A}^{mix}_{CP}$  in
penguin-dominated  $B_s\to K^-K^+,K^-K^{*+}$, $K^{*-}K^+$ and
$\mathcal{B}$, $\mathcal{A}^{L,dir}_{CP}$,
$\mathcal{A}^{L,mix}_{CP}$ , $f_L$ in penguin-dominated  $B_s\to
K^{*-}K^{*+}$ decays. The constrained
$|\lambda''^*_{131}\lambda''_{121}|$-$\phi_{RPV}$ plane shows the
allowed range of $\lambda''^*_{131}\lambda''_{121}$ as shown in
Fig. \ref{rpvbounds}(a). The  $\mathcal{B}(B_s\to K^{*-}K^{+}$,
$K^-K^{*+}$, $K^{*-}K^{*+})$ shown in Fig. \ref{131121}(a-c) have
the similar change trends with
$|\lambda''^*_{131}\lambda''_{121}|$ and $|\phi_{RPV}|$, and they
all increases  with $|\lambda''^*_{131}\lambda''_{121}|$ and
decreases with $|\phi_{RPV}|$.
For the $\mathcal{A}^{dir}_{CP}$/$\mathcal{A}^{L,dir}_{CP}$ shown
in Fig. \ref{131121}(d-g), $\lambda''^*_{131}\lambda''_{121}$
coupling contribution could be significant.
 $|\mathcal{A}^{dir}_{CP}(B_s\to K^{-}K^+)|$
increases when  $|\lambda''^*_{131}\lambda''_{121}|$ is small,
then $\mathcal{A}^{dir}_{CP}(B_s\to K^{-}K^+)$  decreases and its
sign is flipped. $\mathcal{A}^{dir}_{CP}(B_s(\overline{B}_s)\to
K^{*-}K^+)$, $\mathcal{A}^{dir}_{CP}(B_s(\overline{B}_s)\to
K^{-}K^{*+})$ and $\mathcal{A}^{L,dir}_{CP}(B_s\to K^{*-}K^{*+})$
could have smaller range with $|\lambda''^*_{131}\lambda''_{121}|$.
$|\mathcal{A}^{dir}_{CP}(B_s\to K^{-}K^{+},K^{*-}K^{+})|$
 and $|\mathcal{A}^{L,dir}_{CP}(B_s\to
K^{*-}K^{*+})|$
 decrease with $|\phi_{RPV}|$.
$\mathcal{A}^{dir}_{CP}(B_s\to K^{*-}K^+, K^{-}K^{*+})$ and
$\mathcal{A}^{L,dir}_{CP}(B_s\to K^{*-}K^{*+})$ could be  close to
zero  in entire $\phi_{RPV}$ range.
As shown in Fig. \ref{131121}(h-k), four mixing-induced CP
asymmetries are very sensitive to $|\phi_{RPV}|$ but not sensitive
 to $|\lambda''^*_{131}\lambda''_{121}|$.
 For the penguin dominated  process $B_s\to
K^{*-}K^{*+}$,  its longitudinal polarization could be  small as
shown in Fig. \ref{131121}(l), however,   most points  of
$f_L(B_s\rightarrow K^{*-}K^{*+})$  fill  in [0.7,0.9].

\item The effects of $\lambda'^*_{i13}\lambda'_{i12}$ on
$\mathcal{B}$, $\mathcal{A}^{dir}_{CP}$ and $\mathcal{A}^{mix}_{CP}$
of $B_s\to K^-K^+, K^{*-}K^{+}, K^{-}K^{*+}$
 are presented in Fig. \ref{i13i12}. The
constrained $|\lambda'^*_{i23}\lambda'_{i12}|$-$\phi_{RPV}$ plane
is the same as Fig. \ref{rpvbounds}(b). Fig. \ref{i13i12}(a) show
that $\mathcal{B}(B_s\to K^{*-}K^{+})$ increases with
$|\lambda'^*_{i13}\lambda'_{i12}|$ and decreases with
$|\phi_{RPV}|$.
As shown in Fig. \ref{i13i12}(b), at first
$|\mathcal{A}^{dir}_{CP}(B_s\to K^{-}K^+)|$  increase with
$|\lambda'^*_{i13}\lambda'_{i12}|$, then
$\mathcal{A}^{dir}_{CP}(B_s\to K^{-}K^+)$ could occupy the entire
range $[-0.25,0.23]$ when $|\lambda'^*_{i13}\lambda'_{i12}|$ lies
in $[1.6,2.8]\times10^{-3}$, and $|\mathcal{A}^{dir}_{CP}(B_s\to
K^{-}K^+)|$  decreases with $|\phi_{RPV}|$.
$\mathcal{A}^{dir}_{CP}(B_s\to K^{*-}K^+)$ has narrow ranges with
the constrained  $|\lambda'^*_{i13}\lambda'_{i12}|$ and
$|\phi_{RPV}|$.
Fig. \ref{i13i12}(d-f) show the RPV effects in relevant
mixing-induced CPA. $|\mathcal{A}^{mix}_{CP}(B_s\to K^{-}K^+)|$ is
not  sensitive to $|\lambda'^*_{i13}\lambda'_{i12}|$ but sensitive
to $\phi_{RPV}$. $|\mathcal{A}^{mix}_{CP}(B_s\&\overline{B}_s\to
K^{*-}K^+)|$ and $|\mathcal{A}^{mix}_{CP}(B_s\&\overline{B}_s\to
K^{-}K^{*+})|$ decrease with $|\lambda'^*_{i13}\lambda'_{i12}|$ and
they first increase and then decrease with $|\phi_{RPV}|$.

\item In Fig. \ref{112132}, we plot $\mathcal{B}$,
$\mathcal{A}^{dir}_{CP}$ of $B_s\to K^{*-}\pi^+, $ $K^-\rho^+,$
and $\mathcal{B}$, $\mathcal{A}^{L,dir}_{CP}$, $f_L$ of $B_s\to
K^{*-}\rho^+$ decays as functions of
$\lambda''^*_{132}\lambda''_{112}$. The constrained
$|\lambda''^*_{132}\lambda''_{112}|$-$\phi_{RPV}$ plane is the
same as  Fig. \ref{rpvbounds}(c). One can find $\mathcal{B}(B_s\to
K^{*-}\pi^+, K^-\rho^+)$  increase  with
 $|\lambda''^*_{132}\lambda''_{112}|$ and  $|\phi_{RPV}|$. $\mathcal{B}(B_s\to
K^{*-}\rho^+)$  first decreases and then increases with
$|\lambda''^*_{132}\lambda''_{112}|$, and it is not very sensitive
to $|\phi_{RPV}|$.
As shown by Fig. \ref{112132}(d-e),  the squark exchange RPV effects on
$\mathcal{A}^{dir}_{CP}(B\to K^{*-}\pi^+,K^-\rho^+)$ are very small.
 $\mathcal{A}^{dir}_{CP}(B\to K^{*-}\pi^+,K^-\rho^+)$
 first decrease and then increase with
$|\lambda''^*_{132}\lambda''_{112}|$, and they both  increase with
$\phi_{RPV}$. $\mathcal{A}^{L,dir}_{CP}(B\to K^{*-}\rho^+)$ is
sensitive to $\lambda''^*_{132}\lambda''_{112}$ coupling, and
could be enhanced to $\sim 70\%$  when
$|\lambda''^*_{132}\lambda''_{112}|$ is around $3\times10^{-3}$.
$\mathcal{A}^{L,dir}_{CP}(B\to K^{*-}\rho^+)$ first increases
  and then decreases  with $|\lambda''^*_{132}\lambda''_{112}|$, but is not
   sensitive to $\phi_{RPV}$.
$f_L(B\rightarrow K^{*-}\rho^+)$ has the largest allowed range
when $|\lambda''^*_{132}\lambda''_{112}|$ is around
$3\times10^{-3}$. The $\lambda''^*_{132}\lambda''_{112}$ couplings
could decrease $f_L(B\rightarrow K^{*-}\rho^+)$ to 0.42.

\item  Fig. \ref{i13i11} shows the effects of the RPV couplings
$\lambda'^*_{i13}\lambda'_{i11}$ in $B_s\to K^{*-}\pi^+$ decay.
$\mathcal{B}(B_s\to K^{*-}\pi^+)$  increases with
$|\lambda'^*_{i13}\lambda'_{i11}|$ and is insensitive to
$\phi_{RPV}$. $\mathcal{A}^{dir}_{CP}(B\to K^{*-}\pi^+)$ decreases
with $|\lambda'^*_{i13}\lambda'_{i11}|$ and increases with
$\phi_{RPV}$.

\end{itemize}

\section{Conclusions}
In conclusion,   we have studied the eight decay modes $B_s \to K^{(*)-}K^{(*)+}$,
$K^{(*)-}\pi^{+}$, $K^{(*)-}\rho^{+}$  in the RPV SUSY with the QCDF for the hadronic dynamics.
 With the recent experimental data of $B_s$ decays,  we have obtained fairly
 constrained parameter spaces of the RPV couplings. Furthermore,
 using the constrained parameter spaces, we
have shown the RPV SUSY expectations for the other quantities in
$B_s \to K^{(*)-}K^{(*)+}$, $K^{(*)-}\pi^{+}$, $K^{(*)-}\rho^{+}$
decays which have not been measured yet.

We have found that the RPV couplings
$\lambda''^*_{131}\lambda''_{121}$ and
$\lambda'^*_{i13}\lambda'_{i12}$ could significantly affect
penguin-dominated $B_s\to K^{(*)-}K^{(*)+}$ decays.   Within the
parameter spaces already highly constrained  by $B_{s}\to
K^{-}K^{+}$,
 the branching ratios of $B_s\to K^{-}K^{*+}$, $K^{*-}K^{+}$ and $K^{*-}K^{*+}$ could
be enhanced by few times, and the direct CPA and the mixing-induced
CPA are  in quite large ranges. Interestingly,   the longitudinal
polarization fraction of $B_s\to K^{*-}K^{*+}$ could be suppressed
 as low as 0.30. Therefore future experimental measurements of
these decays  could shrink  or reveal
 the relevant NP parameter spaces.
It is found that the squark exchange coupling
$\lambda''^*_{132}\lambda''_{112}$ could have large contributions to
the branching ratios of $B_s \to K^{*-}\pi^{+}$, $K^{-}\rho^{+}$,
and  enhance the longitudinal direct CP asymmetry of $B_s \to
K^{*-}\rho^{+}$ to $\sim 70\%$.
 The longitudinal polarization fraction of $B_s\to K^{*-}\rho^+$ could be suppressed too.
 The slepton exchange coupling $\lambda'^*_{i13}\lambda'_{i11}$ could enhance
the branching ratio of $B_s\to K^{*-}\pi^+$ by few times. We also have
presented correlations between these physical observable quantities
and the constrained parameter spaces of RPV couplings in Figs.
\ref{131121}-\ref{i13i11}. The results in this paper could be useful
for probing RPV SUSY effects and searching direct RPV signals at
Tevatron and LHC in the near future.

\section*{Acknowledgments}
 The work is supported  by National Science
Foundation under contract Nos.10675039 and 10735080.
The work of Ru-Min Wang was
supported by Brain Korea 21 Project.

\end{document}